\documentclass{article}

\usepackage{amssymb}
\usepackage{verbatim}

\newcommand\m{\mathop}
\newcommand\ds{\displaystyle}

\def\le{\left}
\def\ri{\right}
\def\pa{\partial}
\def\1{\mathbf 1}

\renewcommand\L {\Lambda}
\newcommand\un{\underline}

\newtheorem{remark}{Remark}[section]

\newcommand\encadremath[1]{\vbox{\hrule\hbox{\vrule\kern8pt
\vbox{\kern8pt \hbox{$\displaystyle #1$}\kern8pt}
\kern8pt\vrule}\hrule}}
\def\enca#1{\vbox{\hrule\hbox{
\vrule\kern8pt\vbox{\kern8pt \hbox{$\displaystyle #1$}
\kern8pt} \kern8pt\vrule}\hrule}}

\newcommand\figureframey[3]{
\begin{figure}[bth]
\hrule\hbox{\vrule\kern8pt
\vbox{\kern8pt \vbox{
\begin{center}
{\mbox{\epsfysize=#1.truecm\epsfbox{#2}}}
\end{center}
\caption{#3}
}\kern8pt}
\kern8pt\vrule}\hrule
\end{figure}
}

\def\Amat{\m{\mathbb A}}
\def\Bmat{\m{\mathbb B}}

\newcommand{\beq}{\begin{equation}}
\newcommand{\eeq}{\end{equation}}
\newcommand{\bea}{\begin{eqnarray}}
\newcommand{\eea}{\end{eqnarray}}

\renewcommand{\and}{{\qquad {\rm and} \qquad}}

\newcommand{\virg}{{\,\, , \quad}}
\newtheorem{lemma}{Lemma}[section]

 \newcommand{\Tr}{{\,\rm Tr}\:}
\newcommand{\tr}{{\,\rm tr}\:}

\newcommand{\td}[1]{{\widetilde{#1}}}

\newcommand{\e}{{\,\rm e}\,}
\newcommand{\ee}[1]{{{\rm e}^{#1}}}

\newcommand{\D}{{{\hbox{d}}}}

\renewcommand{\a}{\alpha}
\renewcommand{\b}{\beta}

\begin{document}

\pagestyle{empty}
\hfill SPT-03/139
\addtolength{\baselineskip}{0.20\baselineskip}
\begin{center}

\begin{Large}\fontfamily{cmss}
\fontsize{17pt}{27pt}
\selectfont
\textbf{The PDEs of biorthogonal polynomials\\[8pt] arising in the
    two--matrix model
}\footnote{
M.B. was  supported in part by the Natural Sciences and Engineering Research 
Council
of Canada (NSERC) Grant. no. 261229-03,
B.E. was  supported in part by the EC IHP Network ``Discrete geometries: 
from solid state physics to quantum gravity'', HPRN-CT-1999-000161.}
\end{Large}\\
\vspace{1.0cm}
\begin{large} {M.
Bertola}$^{\dagger\ddagger}$\footnote{bertola@crm.umontreal.ca}, 
 { B. Eynard}$^{\dagger
\star}$\footnote{eynard@spht.saclay.cea.fr}
\end{large}
\\
\bigskip
$^{\dagger}$ {\em Centre de recherches math\'ematiques,
Universit\'e de Montr\'eal\\ C.~P.~6128, succ. centre ville, Montr\'eal,
Qu\'ebec, Canada H3C 3J7} \\
\smallskip
$^{\ddagger}$ {\em Department of Mathematics and
Statistics, Concordia University\\ 7141 Sherbrooke W., Montr\'eal, Qu\'ebec,
Canada H4B 1R6} \\ 
\smallskip
$^{\star}$ {\em Service de Physique Th\'eorique, CEA/Saclay \\ Orme des
Merisiers F-91191 Gif-sur-Yvette Cedex, FRANCE } \\
\bigskip
\bigskip

\end{center}

\vspace{20pt}
\begin{center}
{\bf Abstract:}
\end{center}
%
The two-matrix model can be solved by introducing bi-orthogonal
polynomials. In the case the potentials in the measure are
polynomials, finite sequences of bi-orthogonal polynomials (called
{\em windows})
satisfy polynomial  ODEs as well as deformation equations (PDEs) and finite
difference equations ($\Delta$E) which are all Frobenius compatible
and define discrete and continuous isomonodromic deformations for the
irregular ODE, as shown in previous works of ours.

In the one matrix model an explicit and  concise expression for the
coefficients of these systems is known and it allows to relate the partition
function with  the isomonodromic tau-function of the overdetermined system.
Here, we provide the generalization of those expressions to the case of bi-orthogonal
polynomials, which enables us to compute the determinant of the
fundamental solution of the overdetermined system of ODE+PDEs+$\Delta$E.

\newpage
\pagestyle{plain}
\setcounter{page}{1}


\section{Introduction}
The two-matrix model and biorthogonal polynomials have recently
witnessed a renewed interest due to the hope that the asymptotics could be
found from a Riemann--Hilbert approach similar to that used in one-matrix models
(\cite{BlIt} \cite{dkmvz,dkmvz2}, \cite{FIK}, \cite{BEHiso}
\cite{Asym}, \cite{kapa}). Our recent papers have shown some unexpectedly rich algebraic
structure associated to the biorthogonal polynomials for two
polynomial potentials  \cite{BEH, Needs, Asym}.
We recall that the two matrix-model is defined as the ensemble of
pairs of hermitian matrices of size $N$, with the measure:
\beq
{1\over \mathcal Z_N} \ee{-{1\over \hbar}
 \left[ \tr V_1(M_1)+V_2(M_2)-  M_1 M_2 \right] } \D{M_1} \D{M_2}
\eeq
where $\D{M_1} \D{M_2}$ is the standard Lebesgue measure (product of
Lebesgue measures of all real components of $M_1$ and $M_2$),
$V_1$ and $V_2$ are two polynomials (called the potentials) of degree $d_1+1$ and $d_2+1$:
\beq
V_1(x) = \sum_{k=1}^{d_1+1} {u_k\over k} x^k
\virg
V_2(y) = \sum_{k=1}^{d_2+1} {v_k\over k} y^k
\eeq
and $\mathcal Z_N$ is the partition function:
\beq
\mathcal Z_N = \int \ee{-{1\over \hbar}
 \left[ \tr V_1(M_1)+V_2(M_2)- M_1 M_2 \right]} \D{M_1} \D{M_2}
\eeq
The two matrix model has a wide range of applications in physics, in
particular Euclidean 2d gravity i.e. the statistical
physics of a random surface \cite{Kazakov, KazakoVDK, ZJDFG}.

The 2-matrix model can be solved with the help of two families of
bi-orthogonal polynomials
\beq
\pi_n(x) = x^n + \dots
\virg
\sigma_n(y) = y^n + \dots
\eeq
which are such that:
\beq\label{biorthopairing}
\int\int \D{x}\D{y}\,\, \pi_n(x) \sigma_m(y) \,\,\ee{-{1\over \hbar}
[ V_1(x)+V_2(y)- xy] } = h_n \delta_{nm}\ .
\eeq

It was shown in \cite{BEH},
 that a sequence of $d_2+1$ consecutive $\pi_n$'s (or $d_1+1$
consecutive $\sigma_n$'s) obeys a closed ODE with respect to the
``spectral parameter'' $x$ (or $y$)
\beq
-\hbar \frac {\rm d}{{\rm d}x} \pmatrix{\psi_{N-d_2}(x)\cr\vdots\cr
 \psi_N(x)} = \m{D_1}_N(x)  \pmatrix{\psi_{N-d_2}(x)\cr\vdots\cr
 \psi_N(x)}\ ,
\eeq
 as well as deformation equation (PDEs) with respect to infinitesimal
variations of the coefficients of the potentials $V_1, \ V_2$ and
 difference equations with respect to the choice of the consecutive
 elements ($\Delta$E).
The aim of this article is to provide for  an explicit expression of
said ODE and PDEs, which is
 similar in spirit to the work done in \cite{bauldry,bonan} (the
 $\Delta$E are already very simple and can be found in \cite{BEH}).

Here we follow the same logical lines that were followed in our
previous \cite{BEHiso}.\par
We should remark that an explicit expressions for the ODE is given in \cite{BEH}
 but is not convenient for practical purposes.
 For instance it is not at all obvious  how to compute the trace of
the matrix $\ds\m{D_1}_N(x)$ from the expression written in
\cite{BEH}.
We expect that all  the spectral invariants
of $D_1(x)$ will play an essential r\^ole in the  future developments
on the basis that in the one-matrix model \cite{BlIt, BEHiso} the corresponding invariants
allow to put in direct relation the partition function of the model
with the isomonodromic tau-function in the sense of \cite{FIK, JMU,
  JM}. Indeed it is well
known that the one-matrix model is a subcase of the two-matrix model
where one of the potential is Gaussian.

The expressions contained in this paper allow us to easily compute the
first invariants (traces) of the ODE's and PDE's, thus allowing us to evaluate the
determinant of a fundamental system of solutions of the Frobenius
compatible system. Hopefully they will prove useful for the higher
invariants as well.

\begin{remark}[Notation]\label{notation}
In the following we will often consider semi-infinite matrices.
If $M$ is a semi-infinite matrix we denote:
\begin{itemize}
\item $M_+$ the strictly upper triangular part of $M$,
\item $M_-$ the strictly lower triangular part of $M$,
\item $M_k$ the $k^{\rm th}$ diagonal above the main diagonal if $k\geq 0$,
 and the $|k|^{\rm th}$ diagonal below the main diagonal if $k\leq 0$,
\item $M_{+0}=M_+ + {1\over 2} M_0$,
\item $M_{-0}=M_- + {1\over 2} M_0$,
\item $M_{\geq}$ the upper triangular part (including the main diagonal),
\item $M_{\leq}$ the lower triangular part (including the main diagonal).
\end{itemize}

We introduce the following semi--infinite matrices:\par
$\Lambda$ is the shift matrix:
\beq\label{defLambda}
\Lambda_{ij} = \delta_{i+1,j}\ .
\eeq

$\ds \m \Pi_N$ is the projector on the subspace spanned by the first $N+1$ basis vectors:
\beq\label{defPiN}
\m\Pi_N = {\rm diag}(\overbrace{1,1,\dots,1}^{\hbox{$N+1$-times}},0,\dots)\ .
\eeq

We have the simple properties ($\L$ has only a right inverse):
\bea\label{Lambdaproperties}
&&\Lambda \m\Pi_N = \m\Pi_{N-1} \Lambda
\virg
\m \Pi_N \Lambda^t = \Lambda^t \m\Pi_{N-1}\\
&&\Lambda^t \Lambda =\1-\m\Pi_0
\virg
\Lambda \Lambda^t =\1
\virg \Lambda \m\Pi_0=0
\virg \m\Pi_0\Lambda^t=0\ .\label{simpleprop}
\eea

Finally the notation for the folded matrices is slightly changed from
\cite{BEH} in order to make it more consistent (in particular the
location of certain over/under-scripts).
\end{remark}

\section{Biorthogonal polynomials: definitions and properties}
It is convenient to introduce the wave functions and their
Fourier--Laplace transforms, related to the
bi-orthogonal polynomials (\ref{biorthopairing}) by the relations:
\bea
\psi_n(x) = {1\over \sqrt{h_n}} \pi_n(x) \ee{-{1\over \hbar} V_1(x)}
\virg
\phi_n(y) = {1\over \sqrt{h_n}} \sigma_n(y) \ee{-{1\over \hbar}
  V_2(y)}\\
{\un\psi}_n(y) = \int \D{x} \ee{{1\over \hbar} xy} \psi_n(x)
\virg
{\un\phi}_n(x) = \int \D{y} \ee{{1\over \hbar} xy} \phi_n(y)\
.\label{FLT}
\eea
The unspecified contour of integration can be chosen in $d_1$
homologically distinct ways for the functions ${\un\psi}_n(y)$ ($d_2$
ways for ${\un\phi}_n(x)$) as explained in \cite{Asym, Marco}.
We introduce also the semi-infinite wave-vectors
\beq
\mathop{\Psi}_\infty = (\psi_0,\psi_1,\dots)^t
\virg
\mathop{\Phi}_\infty = (\phi_0,\phi_1,\dots)^t
\virg
\mathop{\un\Psi}_\infty = ({\un\psi}_0,{\un\psi}_1,\dots)
\virg
\mathop{\un\Phi}_\infty = ({\un\phi}_0,{\un\phi}_1,\dots)\ .
\eeq
In order to express the ODEs that these wave-functions satisfy we
define the following ``windows'' of consecutive
wave-functions of dimension $d_2+1$ (resp. $d_1+1$):
\beq\label{defwindow}
\mathop{\Psi}_N= (\psi_{N-d_2}, \dots,\psi_{N-1},\psi_{N})^t
\virg
\mathop{\Phi}_N= (\phi_{N-d_1}, \dots,\phi_{N-1},\phi_{N})^t
\eeq
\beq\label{defwindowun}
\mathop{\un\Psi}^N =
({\un\psi}_{N-1},{\un\psi}_{N},\dots,{\un\psi}_{N+d_1-1})
\virg
\mathop{\un\Phi}^N =
({\un\phi}_{N-1},{\un\phi}_{N},\dots,{\un\phi}_{N+d_2-1}) \ .
\eeq
The two windows in each of the pairs $\ds \le(\mathop{\Psi}_N,\mathop{\un\Phi}^N\ri)$ and
$\ds\le(\mathop{\Phi}_N, \mathop{\un\Psi}^N\ri)$ (each constituted of
windows of the same size) are called ``dual windows'' as explained in \cite{BEH}.
It is shown in \cite{BEH,Needs,Asym} that --similarly to the ordinary
orthogonal polynomials--
the  semi-infinite wave-vectors obey:
\beq\label{multxyinfty}
x \m{\Psi}_\infty = Q \m{\Psi}_\infty
\virg
y \m{\Phi}_\infty = P^t \m{\Phi}_\infty
\virg
y \m{\un\Psi}_\infty = \m{\un\Psi}_\infty P^t
\virg
x \m{\un\Phi}_\infty(x) = \m{\un\Phi}_\infty Q
\eeq

\beq\label{derivxyinfty}
-\hbar\pa_x \mathop{\Psi}_\infty = P \mathop{\Psi}_\infty
\virg
-\hbar\pa_y \mathop{\Phi}_\infty =  Q^t \mathop{\Phi}_\infty
\virg
\hbar\pa_y \mathop{\un\Psi}_\infty = \mathop{\un\Psi}_\infty Q
\virg
\hbar\pa_y \mathop{\un\Phi}_\infty =  \mathop{\un\Phi}_\infty P^t
\eeq
where the semi-infinite matrices $P$ and $Q$ are of finite band size.
Componentwise the above equations translate into
\beq
x\psi_n(x) = \gamma(n) \psi_{n+1} + \sum_{k=0}^{d_2} \alpha_k(n) \psi_{n-k}(x)
\virg
y\phi_n(y) = \gamma(n) \phi_{n+1} + \sum_{k=0}^{d_1} \beta_k(n) \phi_{n-k}(y)
\eeq
\beq
x{\un\phi}_n(x) = \gamma({n-1}) {\un\phi}_{n-1} + \sum_{k=0}^{d_2} \alpha_k(n+k) {\un\phi}_{n+k}(x)
\virg
y{\un\psi}_n(y) = \gamma({n-1}) {\un\psi}_{n-1} + \sum_{k=0}^{d_1} \beta_k(n+k) {\un\psi}_{n+k}(y)
\eeq
where $\gamma(n) = \sqrt{h_{n+1}\over h_n}$.

Since they implement the multiplication and differentiation by the
spectral parameters $x$ or $y$, the matrices $P$ and $Q$ obey the Heisenberg relation:
\beq\label{Heisenberg}
[P,Q]= \hbar \mathbf 1
\eeq
and they have the properties \cite{BEH,Needs,Asym}:
\beq\label{eqPQV}
P_+ = V'_1(Q)_+ \virg  Q_- = V'_2(P)_-
\eeq
\beq\label{eqPQVzero}
P_0 = V'_1(Q)_0 \virg  Q_0 = V'_2(P)_0
\eeq
\beq\label{eqPVminusone}
P_{n,n-1} = \gamma(n-1) = V'_1(Q)_{n,n-1} - \hbar {n\over \gamma(n-1)}
 \eeq
\beq\label{eqQVplusone}
Q_{n,n+1} = \gamma(n) = V'_2(P)_{n,n+1} -  \hbar {n+1\over \gamma(n)}
\eeq

The relations (\ref{derivxyinfty}) can be ``folded'' \cite{BEH} onto the finite
windows defined in (\ref{defwindow},\ref{defwindowun}) using the recursion relations (\ref{multxyinfty}) so as
 to give finite dimensional ODEs:
\beq\label{derivxy}
-\hbar\pa_x  \mathop{\Psi}_{N} = {\m{{ D}_1}_N}(x)
\mathop{\Psi}_N
\virg
-\hbar\pa_y \mathop{\Phi}_{N} ={\m{{ D}_2}_N}(y)
\mathop{\Phi}_N
\virg
\hbar\pa_y \mathop{\un\Psi}^{N} =  \mathop{\un\Psi}^N
 {\m{{\un{
D}}_2}^N}(y) \virg
\hbar \pa_x \mathop{\un\Phi}^{N} =  \mathop{\un\Phi}^N
 {\m{{\un{D}}_1}^N}(x)\eeq
where ${\ds \m{{ D}_1}_N}(x)$ and ${\ds \m{{\un{ D}}_1}^N}(x)$
(resp. ${\ds \m{{ D}_2}_N}(y)$ and ${\ds \m{{\un{
D}}_2}^N}(y)$) are matrices of dimension $d_2+1$ (resp. $d_1+1$) whose
entries are polynomials in $x$ (resp. $y$) of degree at most $d_1$
(resp. $d_2$).
They enjoy the duality property (\cite{BEH}, Thm 4.2):
\beq\label{duality}
{\m{{\un{D}}_1}^N}(x) \Amat^N = \Amat^N {\m{{ D}_1}_N}(x)
\virg
{\m{{\un{D}}_2}^N}(y) \Bmat^N = \Bmat^N {\m{{ D}_2}_N}(y)
\eeq
where $\ds{ \Amat^N}$ and $\ds \Bmat^N$ are the Christoffel-Darboux pairing matrices:
\beq\label{defA}
\Amat^N = \pmatrix{
 0 \dots & 0 & -\gamma(N-1) \cr
\alpha_{d_2}(N) \dots & \alpha_1(N) & 0 \cr
\ddots & \vdots & \vdots \cr
0 & \alpha_{d_2}(N+d_2-1) & 0
}
\eeq
\beq\label{defB}
\Bmat^N = \pmatrix{
 0 \dots & 0 & -\gamma(N-1) \cr
\beta_{d_1}(N) \dots & \beta_1(N) & 0 \cr
\ddots & \vdots & \vdots \cr
0 & \beta_{d_1}(N+d_1-1) & 0
}\ .
\eeq
The two C-D pairing matrices $\ds\Amat^N$ and $\ds\Bmat^N$ are the
only non-vanishing blocks in the semi-infinite  matrix commutators:
\beq\label{Acommutator}
[Q,\m{\Pi}_{N-1} ] \qquad ({\rm resp.}\,\,\, [P^t,\m{\Pi}_{N-1}])
\eeq
where $\ds \m{\Pi}_{N-1}$  is the projector defined in (\ref{defPiN}).
These matrices play an essential r\^ole in the computation of the 
relevant spectral statistics of the model inasmuch as 
they enter directly in the kernels used 
in the Dyson--like formulas \cite{eynardmehta}
\bea
(x'-x)\, K_{11}(x,x') = \m{\un\Phi}^N(x')\Amat^N \m{\Psi}_N(x) =
\m{\un\Phi}_\infty(x') \le[Q,\m\Pi_{N-1}\ri] \m{\Psi}_\infty(x)\ \cr
(y'-y)\, K_{22}(y,y') = \m{\un\Psi}^N(y')\Bmat^N \m{\Phi}_N(y) =
\m{\un\Psi}_\infty(y') \le[Q,\m\Pi_{N-1}\ri] \m{\Phi}_\infty(y)\ .
\eea
\subsection{Deformation equations (PDE's)}
The deformations of the biorthogonal polynomials
(derivations w.r.t the coefficients of the two potentials), are
represented by semi-infinite finite-band matrices:
\beq\label{deforminfty}
\hbar \pa_{u_K} \mathop{\Psi}_\infty = U_K
\mathop{\Psi}_\infty
\virg
\hbar\pa_{v_J} \mathop{\Psi}_\infty = - V_J
\mathop{\Psi}_\infty
\virg
\hbar\pa_{u_K}\mathop{\un\Phi}_\infty = -
\mathop{{\un \Phi}}_\infty  U_K
\virg
\hbar\pa_{v_J} \mathop{\un\Phi}_\infty =
\mathop{{\un \Phi}}_\infty   V_J
\eeq

\beq\label{deforminftybis}
\hbar\pa_{u_K} \mathop{\un\Psi}_\infty =
\mathop{{\un \Psi}}_\infty   {U}_K^t
\virg
\hbar\pa_{v_J} \mathop{\un\Psi}_\infty =-
\mathop{{\un \Psi}}_\infty   {V}_J^t
\virg
\hbar\pa_{u_K}\mathop{\Phi}_\infty = - U_K^t
\mathop{\Phi}_\infty
\virg
\hbar\pa_{v_J} \mathop{\Phi}_\infty =  V_J^t
\mathop{\Phi}_\infty
\eeq
where the semi-infinite matrices $U_K,V_J$ are related to the matrices
$P, Q$ by the equations
\bea
U_K:=-\frac 1 K Q^K_{+0}\ ,\qquad V_J:= -\frac 1 J P^J_{-0}\ .
\eea
which can be folded onto the windows using eq.(\ref{multxyinfty}) as
explained in \cite{BEH}
\beq\label{deform}
\hbar\pa_{u_K} \mathop{\Psi}_{N} =  {\mathop{{\cal U}_K}_{\!\!N}}(x) \mathop{\Psi}_N
\virg
\hbar\pa_{v_J} \mathop{\Psi}_{N} = -{\mathop{{\cal
V}_J}_{\!\!\!\!N}}(x) \mathop{\Psi}_N
\virg
\hbar\pa_{ u_K}  \mathop{\un\Phi}^{N} =  - \mathop{\un\Phi}^N {\mathop{{\un{\cal
U}}}^N}_K(x)
\virg
\hbar\pa_{v_J} \mathop{\un\Phi}^{N} =  \mathop{\un\Phi}^N  {\mathop{{\un{\cal
V}}}^N}_J(x)
\eeq\par
\beq\label{deformbis}
\hbar\pa_{ u_K} \mathop{\un\Psi}^{N} =  \mathop{\un\Psi}^N  {\mathop{\td{\cal
U}}^N}_K(y)
\virg
\hbar\pa_{ v_J} \mathop{\un\Psi}^{N} = - \mathop{\un\Psi}^N {\mathop{\td{\cal
V}}^N}_J(y)
\virg
\hbar\pa_{ u_K}\mathop{\Phi}_{N} = - {\mathop{{{\td{\cal
U}}}}_N}\,\!_K(y) \mathop{\Phi}_N
\virg
\hbar\pa_{v_J} \mathop{\Phi}_{N} =  {\mathop{{{\td{\cal
V}}}}_N}\,\!_J(y) \mathop{\Phi}_N
\eeq
where $\ds{\mathop{{\cal U}}_N}\!\,_K(x)$ and $\ds{\mathop{{\un{\cal
U}}}^N}_K(x)$ are matrices of dimension $d_2+1$  whose entries are
polynomials in $x$ of degree at most $k$,
and $\ds{\mathop{{\cal V}}_N}\!\,_J(x)$ and $\ds{\mathop{{\un{\cal V}}}^N}_J(x)$
are matrices of dimension $d_2+1$  whose entries are polynomials in
$x$ of degree at most $J d_1$ (and similar statement for the
tilde-matrices with $d_1$ replacing $d_2$ and vice-versa).
They satisfy \cite{BEH}:
\bea
&& {\mathop{{\un{\cal U}}}^N}_K(x) \Amat^N + \Amat^N {\mathop{{\cal
U}}_{\!N}\!_K}(x) + \hbar\pa_{u_K}{\ds\Amat^N}=0\\
&& {\mathop{{\un{\cal V}}}^N}_J(x) \Amat^N + \Amat^N {\mathop{{\cal
V}}_{\!N}\!_J}(x) - \hbar\pa_{v_J}{\ds\Amat^N}=0\ .
\eea
The expressions of the matrices $U_J,V_K$ in terms of $P$ and $Q$ can
be found with the same notation in \cite{Asym} but are essentially
well known since \cite{UT} in a slightly different context and with
different notation.

\subsection{Deformations with respect to $\hbar$}

The $\hbar$-deformations of the biorthogonal polynomials (derivations
w.r.t $\hbar$) are also represented by semi-infinite
matrices of finite  band size which we can write as follows:
\beq\label{deformhinfty}
\hbar^2\pa_\hbar \mathop{\Psi}_\infty = H \mathop{\Psi}_\infty
\virg
\hbar^2\pa_\hbar \mathop{\Phi}_\infty =  \widetilde H \mathop{\Phi}_\infty .
\eeq
The relationship with the matrices $P$ and $Q$ is easy to derive but
since we could not find it in the literature (except in special cases \cite{kapa}) 
we give it hereafter.
From $\int \e^{\frac {xy}\hbar} \m{\Psi}_\infty(x)
{\m{\Phi}_\infty}^t(y) =  \1$ we promptly obtain
\beq
H-QP+\widetilde H^t = 0\ .
\eeq
Moreover we have
\bea
&& \le(H-V_1(Q)\ri)_+ = 0  = \le(\widetilde H - V_2(P^t)\ri)_+\\
&&\overbrace{H-V_1(Q)}^{\hbox{lower}}
 + \le(V_1(Q)-QP+V_2(P)\ri) + \overbrace{\widetilde H^t
-V_2(P)}^{\hbox{upper}}=0 \\
&&\le(H-V_1(Q)\ri)_0 =  \le(\widetilde H - V_2(P^t)\ri)_0=-\frac 1
2\le(V_1(Q)-QP+V_2(P)\ri) _0 \\
&& \le(H-V_1(Q)\ri)_- =  \le(QP-V_1(Q)-V_2(P)\ri)_-\\
&& \le(\widetilde H- V_2(P^t)\ri)_-=\le(P^tQ^t-
V_1(Q^t)-V_2(P^t)\ri)_-
\eea
so that we have
\bea\label{HinPQ}
H &=& V_1(Q)_+ + \frac 1 2 (QP+V_1(Q)-V_2(P))_0 + (QP-V_2(P))_- =
V_1(Q)_{+0} + (QP)_{-0} -V_2(P)_{-0}\\
\widetilde H &=&
V_2(P^t)_{+0} + (P^tQ^t)_{-0} -V_1(Q^t)_{-0}\ .
\eea

It is also interesting to consider another derivation of (\ref{HinPQ}), from scaling
properties.
Consider the change of variables $x\mapsto \frac x\lambda$ and $y\mapsto \frac y\mu$, 
it gives
\beq
(\lambda\mu)^{n+1} h_n(\{u_K\},\{v_J\},\hbar) \delta_{nm} = \int {\rm d}x{\rm d}y
\lambda^n\pi_n\le(\frac x \lambda,\{u_K\},\{v_J\},\hbar \ri)
\mu^m\sigma_m\le(\frac y \mu,\{u_K\},\{v_J\},\hbar \ri) {\rm
  e}^{-\frac {\widetilde V_1(x)+\widetilde V_2(y)-xy} {\widetilde\hbar}}
\eeq
where
\bea
\widetilde V_1(x) := \lambda \mu V_1\le(\frac x\lambda\ri)\ ,\qquad
 \widetilde V_2(y) := \lambda \mu V_2\le(\frac y\mu\ri)\ ,\qquad
 \widetilde \hbar = \lambda\mu\hbar\ .
\eea
This immediately implies the identities
\bea
\pi_n(x;\{u_K\lambda^{1-K}\mu\},\{v_J\mu^{1-J}\lambda\},\lambda\mu\hbar)
= \lambda^n\pi_n\le(\frac x \lambda,\{u_K\},\{v_J\},\hbar \ri)\\
\sigma_n(y;\{u_K\lambda^{1-K}\mu\},\{v_J\mu^{1-J}\lambda\},\lambda\mu\hbar)
= \mu^n\sigma_n\le(\frac y \mu,\{u_K\},\{v_J\},\hbar \ri)\\
h_n\le(\{u_K\lambda^{1-K}\mu\},\{v_J\mu^{1-J}\lambda\},\lambda\mu\hbar\ri)
= (\lambda\mu)^{n+1} h_n\le(\{u_K\},\{v_J\},\hbar \ri)\\
\psi_n\le(x,\{u_K\lambda^{1-K}\mu\},\{v_J\mu^{1-J}\lambda\},\lambda\mu\hbar\ri)
= \lambda^{\frac {n-1}2} \mu^{-\frac{n+1}2} \psi_n\le(\frac x \lambda;\{u_K\},\{v_J\},\hbar\ri)\\
\phi_n\le(y,\{u_K\lambda^{1-K}\mu\},\{v_J\mu^{1-J}\lambda\},\lambda\mu\hbar\ri)
= \mu^{\frac {n-1}2} \lambda^{-\frac{n+1}2} \phi_n\le(\frac y \mu;\{u_K\},\{v_J\},\hbar\ri)
\eea

In infinitesimal form, from $\lambda{\pa_\lambda}_{|_{\lambda=\mu=1}}$
and  $\mu{\pa_\mu}_{|_{\lambda=\mu=1}}$ we obtain
\bea
\le(\frac {\widehat n-1} 2 + x\pa_x -\sum_K (K-1)u_K\pa_{u_K} + \sum_J v_J\pa_{v_J}
+ \hbar\pa_\hbar \ri)\m{\Psi}_\infty(x) = 0\\
\le(-\frac {1+\widehat n} 2 + \sum_K u_K\pa_{u_K} - \sum_J (J-1)v_J\pa_{v_J}
+ \hbar\pa_\hbar \ri)\m{\Psi}_\infty(x) = 0\\
\le(-\frac {1+\widehat n} 2  -\sum_K (K-1)u_K\pa_{u_K} + \sum_J v_J\pa_{v_J}
+ \hbar\pa_\hbar \ri)\m{\Phi}_\infty(y) = 0\\
\le(\frac {\widehat n-1} 2  +y\pa_y + \sum_K u_K\pa_{u_K} - \sum_J (J-1)v_J\pa_{v_J}
+ \hbar\pa_\hbar \ri)\m{\Phi}_\infty(y) = 0\ ,
\eea
where $\widehat n := {\rm diag}(0,1,2,3,4,\dots)$.
Using eq. (\ref{deforminfty}) we immediately obtain that
\bea
-H = \frac {\hbar(\widehat n-1)} 2  - PQ- \sum_K(K-1)u_K U_K- \sum_J v_J V_J =
 -\frac {\hbar(\widehat n+1)} 2 + \sum_{K}u_KU_K+\sum_J(J-1)v_JV_J\\
-\widetilde H^t = -\frac{\hbar(\widehat n+1)}2  + \sum (K-1)u_K U_K +
\sum v_JV_J = \frac {\hbar(\widehat n-1)}2 - PQ - \sum u_KU_K -
\sum(J-1)v_JV_J\ ,
\eea
which, using (\ref{eqPVminusone}, \ref{eqQVplusone}), is identical to (\ref{HinPQ}).

We also get the identity
\beq
-PQ = \sum_J J v_J V_J +\sum_K K u_K U_K-\hbar\,\widehat n\ .
\label{PQrel}
\eeq
This last relation is nothing but the expression  of the scaling
properties of the biorthogonal polynomials and will prove itself very useful
later on.

\subsection{Ladder recursion relations ($\Delta$E's)}

We give here only the main statements without proofs and refer the
reader to ref. \cite{BEH} for further details.
Let us introduce the sequence of companion--like  matrices $\ds{\m{\bf
a}_N(x)}$
and $\ds{ \m{\un{\bf a}}^N(x)}$ of size $d_2+1$
\bea
\m{\bf a}_N(x) & := &
\le[\begin{array}{cccc} 0  & 1 & 0 &\!\!\! \!\!\!\!\!\! \!\!\! 0 \cr
0 & 0 & \ddots &\!\!\! \!\!\!\!\!\! \!\!\!0\cr
0 & 0 & 0 &\!\!\! \!\!\!\!\!\! \!\!\!1\cr
\!\!\frac {-\alpha_{d_2}(N)} {\gamma(N)}\!\! &
\cdots
&\!\! \frac {-\alpha_1(N)}
{\gamma(N)} \!\! &\!\!  \frac{(x-\alpha_0(N))}{\gamma(N)} \!\!
\end{array}\ri] \ ,  \quad N\geq d_2 \ ,  \label{aNdef}\\
 \m{\un{\bf a}}^{N}(x) & := & \le[
\begin{array}{cccc}
  \frac{x\!-\!\a_0\!(N)}{\gamma(N\!-\!1)}&1&0&0\\[5pt]
 \frac{-\a_1\!(N\!+\!1)}{\gamma(N\!-\!1)}&0&
^{\ds{^{\ds{\cdot}}}}\hbox{}
 ^{\ds{\cdot}} \cdot&0\\
 \vdots &0&0&1\\[5pt]
   \frac{-\a_{d_2}\!(N\!+\!d_2)}{\gamma(N\!-\!1)}&0&0&0
\end{array} \ri]\ ,\ N\geq 1,
\label{abarN}
\eea
and also the analogous sequence of matrices $\ds{\m{\bf b}_N(y)}$
and $\ds{ \m{\un{\bf b}}^N(y)}$ of size $d_1+1$
\bea
\m{\bf b}_N(y) & := &
\le[\begin{array}{cccc} 0  & 1 & 0 &\!\!\! \!\!\!\!\!\! \!\!\! 0 \cr
0 & 0 & \ddots &\!\!\! \!\!\!\!\!\! \!\!\!0\cr
0 & 0 & 0 &\!\!\! \!\!\!\!\!\! \!\!\!1\cr
\!\!\frac {-\beta_{d_1}(N)} {\gamma(N)}\!\! &
\cdots
&\!\! \frac {-\beta_1(N)}
{\gamma(N)} \!\! &\!\!  \frac{(y-\beta_0(N))}{\gamma(N)} \!\!
\end{array}\ri] \ ,  \quad N\geq d_1 \ ,  \label{bNdef}\\
 \m{\un{\bf b}}^{N}(y) & := & \le[
\begin{array}{cccc}
  \frac{y\!-\!\b_0\!(N)}{\gamma(N\!-\!1)}&1&0&0\\[5pt]
 \frac{-\b_1\!(N\!+\!1)}{\gamma(N\!-\!1)}&0&
^{\ds{^{\ds{\cdot}}}}\hbox{}
 ^{\ds{\cdot}} \cdot&0\\
 \vdots &0&0&1\\[5pt]
   \frac{-\b_{d_1}\!(N\!+\!d_1)}{\gamma(N\!-\!1)}&0&0&0
\end{array} \ri]\ ,\ N\geq 1.
\label{bbarN}
\eea
The equations in (\ref{multxyinfty})  imply  the
following
\begin{lemma} \label{abNrecursions}
The sequences of matrices  $\ds{\m{\bf a}_N}$, $\ds{\m{\bf b}_N}$ and
$\ds{\m{\un{\bf a}}^N}$, $\ds{\m{\un{\bf b}}^N}$ implement the
shift  $N\mapsto N+1$ and $N\mapsto N-1$ in the windows of
quasi-polynomials and Fourier--Laplace transforms in the sense that
\bea
 \m{\bf a}_N\m{\Psi}_N(x) =
\m{\Psi}_{N\!+\!1}(x) \ ,\qquad
\m{\un\Phi}^{N}(x)  =
\m{\un\Phi}^{N\!+\!1}(x)\m{\un{\bf a}}^N \ ,\label{ladder}\\
\m{\bf b}_N\m{\Phi}_N(y) =
\m{\Phi}_{N\!+\!1}(y) \ ,\qquad
\m{\un\Psi}^{N}(y)  =
\m{\un\Psi}^{N\!+\!1}(y)\m{\un{\bf b}}^N \ . \label{ladder2}
\eea
 and in general
\bea
\m{\Psi}_{N\!+\!j} = \m{\bf a}_{N+j-1}\cdots\m{\bf a}_{N}
\m{\Psi}_N\ ,\qquad \m{\un\Phi}^{N}  =
\m{\un\Phi}^{N\!+\!j}\m{\un{\bf a}}^{N+j-1}\cdots \m{\un{\bf a}}^{N}\ ,\label{folding}\\
\m{\Phi}_{N\!+\!j} = \m{\bf b}_{N+j-1}\cdots\m{\bf b}_{N}
\m{\Phi}_N\ ,\qquad \m{\un\Psi}^{N}  =
\m{\un\Psi}^{N\!+\!j}\m{\un{\bf b}}^{N+j-1}\cdots \m{\un{\bf b}}^{N}\ ,\label{folding1}
\eea
where $\ds{\m{\un\Psi}^{N}\ ,\m{\un\Phi}^{N} }$ here denotes a window in any of the
Fourier--Laplace transforms defined in eqs. (\ref{FLT}).
\end{lemma}

It was proven in \cite{BEH} that the
 differential-deformation-recursion relations
 (\ref{derivxy}, \ref{deform}, \ref{ladder}) are compatible and admit a common fundamental solution.

\subsection{Folding}
{\bf Important remark}.
From this point on we focus on the wave-functions $\psi_n(x)$ and
their dual $\un\phi_n(x)$, but
everything being said can be immediately extended to the wave-functions
$\phi_n(y), \un\psi_n(y)$ by interchanging the r\^oles of the spectral parameters,
the potentials and the matrices $P$ and $Q$.\par
We recall that the notion of ``folding'' is the following: to express
any quasipolynomial $\psi_n(x)$ as a linear combination of $d_2+1$
{\em fixed} consecutive quasipolynomials $\psi_{N-j}\ ,\ j=0,\dots
d_2$ {\em with polynomial coefficients}.
We now provide a way of
computing the folding which is different  from (but equivalent to) the one used in
\cite{BEH} where the ladder matrices were used instead.\par
Let $N$ (i.e. a window) be fixed; we seek to describe the folding of the
infinite wave-vector $\ds\m{\Psi}_\infty(x)$ onto the window at $N$ by
means of a single matrix $\ds\m {\mathcal F}_N (x)$ of size $\infty\times(d_2+1)$
 and with polynomial entries such that:
\beq
\forall n,\qquad 
\psi_n(x) = \sum_{k=N-d_2}^{N} {\m{{\mathcal F}}_N}\!\,_{n,k}(x) \psi_{k}(x)\ .
\eeq
In fact it is more convenient to think  of $\ds\m {\mathcal F}_N$ as a $\infty\times\infty$
matrix with only a vertical band of width $d_2+1$ of nonzero entries
(with column index in the range $N-d_2,\dots,  N$).
In order to describe as explicitly as
possible the matrix $\ds\m {\mathcal F}_N$ we first introduce a convenient
notation to express the band-matrices $P,Q$:
\beq
Q= \gamma\Lambda + \sum_{j=0}^{d_2} \alpha_j {\Lambda^t}^j
\virg
P= \Lambda^t \gamma + \sum_{j=0}^{d_1} \Lambda^j \beta_j
\eeq
where $\Lambda$ is the shift matrix defined in (\ref{defLambda}),
and $\alpha_j,\beta_j,\gamma$ here above denote the diagonal matrices
\beq
\alpha_j := {\rm diag}(\alpha_j(0),\dots,\alpha_j(n),\dots)\ ,\qquad
\beta_j := {\rm diag}(\b_j(0),\dots,\b_j(n),\dots)\ ,\qquad
\gamma := {\rm diag}(\gamma(0),\dots,\gamma(n),\dots)\ .
\eeq
It is known that the existence of biorthogonal polynomials satisfying 
(\ref{biorthopairing}) is equivalent to saying that $\gamma(n)\neq 0, \,\, \forall n$.
Therefore we assume that $\gamma$ is invertible.

Eqs. (\ref{eqPQV}) imply that
\beq
\alpha_{d_2} {\Lambda^t}^{d_2} = v_{d_2+1} (\Lambda^t \gamma)^{d_2}\ ,
\eeq
or, more transparently,
\beq
\alpha_{d_2} (n) = v_{d_2+1} \gamma(n-1)\cdots \gamma(n-d_2)\ .
\eeq

Let us now define the following semi-infinite matrices:
\beq
A:= \1-{1\over v_{d_2+1}}(\gamma^{-1}\Lambda)^{d_2}\, (Q-x)
\virg
B:= \1-\m\Pi_0-(\Lambda^t \gamma^{-1})\, (Q-x)\label{Afol}\ .
\eeq
We note that $A$ is strictly upper triangular while $B$ is strictly lower triangular.

When acting on the semi--infinite  wave-vector $\ds
 \mathop{\Psi}_\infty (x)$
by definition we have  $\ds(Q-x)\mathop{\Psi}_\infty =0$ so that:
\beq
\mathop{\Psi}_\infty =A\mathop{\Psi}_\infty
\virg
\mathop{\Psi}_\infty =\le(B+\m\Pi_0\ri)\mathop{\Psi}_\infty
\eeq

Notice that  $\1-A$ and $\1-B$ are invertible because they are upper
(resp. lower) triangular with ones on the main diagonal.
Notice also that the matrix  $Q-x$ has a left inverse\footnote{The fact that $Q-x$ has a left 
inverse $(Q-x)^{-1}_{\rm L}$ is not in contradiction with 
$(Q-x)\mathop{\Psi}_\infty=0$.\par
Indeed, $\left((Q-x)^{-1}_{\rm L}(Q-x)\right)\mathop{\Psi}_\infty
\neq (Q-x)^{-1}_{\rm L}\left((Q-x)\mathop{\Psi}_\infty\right)$
because both LHS and RHS are infinite sums, and the order of summation 
cannot be exchanged.}:
\beq
{1\over v_{d_2+1}}\, (\1-A)^{-1}\, (\gamma^{-1} \Lambda)^{d_2} (Q-x) = \1
\eeq
{\em and} a right inverse (write $\1-B = {\ds\m\Pi_0} + \L^t\gamma^{-1}(Q-x)$,
and multiply by $\gamma\L$ on the left and by 
$(\1-B)^{-1}\L^t\gamma^{-1}$ on the right):
\beq
(Q-x) (\1-B)^{-1}\, (\Lambda^t \gamma^{-1}) = \1
\eeq
but they do not coincide (one is upper and the other is lower
triangular).

\subsubsection{Upper folding}
We start our construction of the folding matrix $\ds\m {\mathcal F}_N$ by first
looking at the ``upper'' folding, i.e. the part of $\ds\m {\mathcal F}_N$ with
first index greater than $N$.\par
We introduce the following notations
\beq
\m\Pi^N:= \mathbf 1 - \m \Pi_N\ ,\qquad \m\Pi_N^M := \m\Pi_N-\m\Pi_M =
\m\Pi^M-\m\Pi^N = \m\Pi_N\m\Pi^M\ ,
\eeq
and we remark the following formulas which hold since $B$ is
strictly lower triangular
\bea
\m\Pi_N B \m\Pi^N =0\ ;\qquad \m\Pi^N B \m \Pi^N = B\m\Pi^N.
\eea
With these formulas in mind we compute
\bea
&&\m\Pi^N \m\Psi_\infty =\m\Pi^N B \m\Psi_\infty = \m\Pi^N B \le(\m\Pi_N
+ \m\Pi^N\ri)\m\Psi_\infty =
 \m\Pi^N B \m\Pi_N \m\Psi_\infty +  \m\Pi^N B\m\Pi^N\m\Psi_\infty
=   \m\Pi^N B \m\Pi_N \m\Psi_\infty
+  \m\Pi^N B
\le(\m\Pi^N B \m\Pi_N \m\Psi_\infty +  \m\Pi^N B\m\Pi^N\m\Psi_\infty
\ri)=\cr
&& =\le(\mathbf 1 +B\ri) \m\Pi^N B \m\Pi_N \m\Psi_\infty  +\m\Pi^N B
\m\Pi^N B\m\Pi^N\m\Psi_\infty =\le(\mathbf 1 +B\ri) \m\Pi^N B \m\Pi_N
\m\Psi_\infty  +B^2\m\Pi^N\m\Psi_\infty =\hbox{[iterating
    $r$-times]}=\cr
&&=\sum_{j=0}^r B^j \m\Pi^N B \m\Pi_N
\m\Psi_\infty  + B^{r+1}\m\Pi^N\m\Psi_\infty\ .
\eea
Given that $B$ is lower triangular (in a sloppy sense it is nilpotent)
the remainder in the formula above is a vector whose nonzero
components are only those with index greater than $N+r+1$. Since in
our problems we are only folding finite band matrices and we will
never need arbitrarily high indices, we can disregard the remainder
and send $r\to\infty$ (i.e. we take an  inductive limit).
The result is then
\beq
\m\Pi^N \m\Psi_\infty =(1-B)^{-1} \m\Pi^N B \m\Pi_N
\m\Psi_\infty\ .\label{123}
\eeq
Finally note that, since the matrix $B$ has only $d_2$ bands below the
main diagonal, we have
\beq
 \m\Pi^N B \m\Pi_N \m\Pi_{N-d_2}^N = 0\ .
\eeq
In other words the expression in eq.(\ref{123}) contains only the
quasipolynomials $\psi_{N-d_2},\dots,\psi_N$. Therefore we have
achieved the first part of our computation
\beq
\m\Pi^N\, \m {\mathcal F}_N= (1-B)^{-1}\,\m\Pi^N B\, \m \Pi_N\ ,
\eeq
which can be simplified further as follows
\bea
\m\Pi^N\, \m {\mathcal F}_N
& =  & (\1-B)^{-1}\,\m\Pi^N\, B\, \m\Pi_N\,=   \cr
& =  & -(\1-B)^{-1}\, \m\Pi^N\, (\Lambda^t \gamma^{-1})\, (Q-x)\, \m\Pi_N\, =  \cr
& =  & -(\1-B)^{-1}\, (\Lambda^t \gamma^{-1})\, \m\Pi^{N-1}\, (Q-x)\, \m\Pi_N\, =  \cr
& =  & -(\1-B)^{-1}\, (\Lambda^t \gamma^{-1})\, (Q-x)\, \m\Pi^{N-1}\,
\m\Pi_N\,  -(1-B)^{-1}\, (\Lambda^t \gamma^{-1})\, \le[Q-x,
  \m\Pi_{N-1}\ri]\,\m \Pi_N\, =  \cr
& =  &- (\1-B)^{-1}\, \le(\1-B-\m\Pi_0\ri)\, \m\Pi_N^{N-1} -(\1-B)^{-1}\, (\Lambda^t
\gamma^{-1})\, \le[Q, \m\Pi_{N-1}\ri]\, = \cr
& =  &- \m\Pi_N^{N-1}\,
       -(\1-B)^{-1}\, (\Lambda^t \gamma^{-1})\, \le[Q, \m\Pi_{N-1}\ri]\, \cr
\eea
Concluding this part we have
\beq
\m\Pi^N \m {\mathcal F}_N =- \m\Pi_N^{N-1}\,
 -(1-B)^{-1}\, (\Lambda^t \gamma^{-1})\, \le[Q,
   \m\Pi_{N-1}\ri]\label{upperfolding}\ .
\eeq
\subsubsection{Lower folding}
Similarly to the case before we have now
\beq
\m\Pi^{N-d_2-1} A\m\Pi_{N-d_2-1} =0\ ; \ \ \m\Pi_{N-d_2-1} A\m\Pi_{N-d_2-1} = A\m\Pi_{N-d_2-1}
\eeq
and hence we promptly obtain along the same lines (we set for
convenience $M=N-d_2-1$)
\bea
&& \m\Pi_{M}\m\Psi_\infty =
\m\Pi_{M}A\m\Psi_\infty = \m\Pi_{M}A\m\Pi^{M}\m\Psi_\infty+
\m\Pi_{M}A\m\Pi_{M}\m\Psi_\infty =\cr
&&= \m\Pi_{M}A\m\Pi^{M}\m\Psi_\infty+
\m\Pi_{M}A\le( \m\Pi_{M}A\m\Pi^{M}\m\Psi_\infty+
\m\Pi_{M}A\m\Pi_{M}\m\Psi_\infty\ri) = \cr
&&=
\le(1+A\ri)\m\Pi_{M}A\m\Pi^{M}\m\Psi_\infty
+ A^2\m\Pi_{M}\m\Psi_\infty = \hbox{ [iterating $r$-times] }=\cr
&&=\sum_{j=0}^r A^j
\m\Pi_{M}A\m\Pi^{M}\m\Psi_\infty +A^{r+1}
\m\Pi_{M}\m\Psi_\infty\ .
\eea
For the same reason used above we disregard the remainder term and
thus we obtain
\beq
 \m\Pi_{M}\m\Psi_\infty= (1-A)^{-1}
 \m\Pi_{M}A\m\Pi^{M}\m\Psi_\infty\ .
\eeq
Once more this expression contains only the quasipolynomials within
the window at $N$.
Therefore:
\beq
\m\Pi_{M} \m {\mathcal F}_N =  (1-A)^{-1}\, \m\Pi_{M}\, A\, \m\Pi^{M}\,
\eeq
Once and again this expression can be simplified further as follows
\bea
\m\Pi_{M} \m {\mathcal F}_N
& =  & (1-A)^{-1}\, \m\Pi_{M}\, A\, \m\Pi^{M}\, =   \cr
& = & - {1\over v_{d_2+1}}\, (1-A)^{-1}\, \m\Pi_{M}\,
(\gamma^{-1} \Lambda)^{d_2}\, (Q-x)\, \m\Pi^{M}=    \cr
& = & - {1\over v_{d_2+1}}\, (1-A)^{-1}\, (\gamma^{-1}
\Lambda)^{d_2}\, \m\Pi_{N-1}\, (Q-x)\,\m\Pi^{M}\,  =   \cr
& = & - {1\over v_{d_2+1}}\, (1-A)^{-1}\, (\gamma^{-1}
\Lambda)^{d_2}\, (Q-x)\, \m\Pi_{N-1}\,\m\Pi^{M}\,  \cr
&   & + {1\over v_{d_2+1}}\, (1-A)^{-1}\, (\gamma^{-1}
\Lambda)^{d_2}\, [Q-x, \m\Pi_{N-1}]\,=      \cr
& = &  -(1-A)^{-1}\, (1-A)\, \m\Pi_{N-1}^{M}\,    \cr
&   & + {1\over v_{d_2+1}}\, (1-A)^{-1}\, (\gamma^{-1}
\Lambda)^{d_2}\, [Q, \m\Pi_{N-1}]\,=      \cr
& = &- \m\Pi_{N-1}^{M}\,
      + {1\over v_{d_2+1}}\, (1-A)^{-1}\, (\gamma^{-1}
      \Lambda)^{d_2}\, [Q, \m\Pi_{N-1}]\ .\label{lowerfolding}
\eea
Finally we remark that $\ds \m {\mathcal F}_N\,\!_{i,j}$ for $N-d_2\leq i,j\leq N$ is the
$d_2+1$ identity matrix on the selected window so that, putting
eqs. (\ref{upperfolding}, \ref{lowerfolding})  together we have:
\bea
\m{\mathcal F}_N
& = & \m\Pi_{M} \m{\mathcal F}_N + \m\Pi^{N} \m{\mathcal F}_N  + \m\Pi_{N}^{M} \cr
& = & -(1-B)^{-1}\, (\Lambda^t \gamma^{-1})\, \le[Q, \m\Pi_{N-1}\ri]  +
{1\over v_{d_2+1}}\, (1-A)^{-1}\, (\gamma^{-1} \Lambda)^{d_2}\, \le[Q,
  \m\Pi_{N-1}\ri]\, \cr
\eea
We can summarize the above computations into the single formula
\beq
\encadremath{
\m {\mathcal F}_N = \left( {1\over v_{d_2+1}}\, (1-A)^{-1}\, (\gamma^{-1} \Lambda)^{d_2} -(1-B)^{-1}\, (\Lambda^t \gamma^{-1})\right)\,
 \le[Q, \m\Pi_{N-1}\ri]
}\eeq
Using this folding operator we can immediately find the folded version
of any other operator, viz,
if $L$ is any semi-infinite matrix with finite band size, its folded counterpart is:
\beq
L\,\, \longrightarrow \,\,  {\cal L}(x) = \m\Pi_{N}^M L\,
{\mathcal F}_N\label{opfolding}\ .
\eeq
Note that --strictly speaking-- formula (\ref{opfolding}) defines a
semiinfinite matrix with only a nonzero square block of size $d_2+1$
located on the main diagonal
\beq
 {\cal L}(x)_{ij}\equiv 0\ ,i,j\not \in {N-d_2,\dots,N}\ .
\eeq
By abuse of notation we will denote by the same symbol the
$(d_2+1)\times (d_2+1)$ matrix corresponding to the nonzero block. In
this fashion we will not distinguish between tracing over the finite
or the semiinfinite matrix (with some advantages which will appear
later).
\subsection{Dual Folding}
For completeness we report the formulas for the folding of the dual
wave vector $\ds\m{\un\Phi}_\infty$ onto the window $N-1\leq n\leq
N+d_2-1$. The computation is completely parallel to the  above using now
the matrices (remember that the dual wave vector is a row-vector and
hence we must act on the right)
\beq
{\un B} = \1- (Q-x)\L^t\gamma^{-1}\ ,\qquad {\un A} = \1-\m\Pi_{d_2-1} -
(Q-x) \frac{(\gamma^{-1}\L)^{d_2}}{v_{d_2+1}}\ .
\eeq
With this in mind it is straightforward to obtain
\bea
 &&\m{\un\Phi}_\infty\m\Pi_{N-2} =\m{\un \Phi}_\infty\m\Pi^{N-2} {\un B} \m\Pi_{N-2}
(\1-{\un B})^{-1} = -\m\Pi_{N-1}^{N-2} - \le[Q,\m\Pi_{N-1}\ri]
\L^t\gamma^{-1}(\1-{\un B})^{-1}\cr
&&  \m{\un\Phi}_\infty\m\Pi^{N+d_2-1} =\m{\un\Phi}_\infty\m\Pi_{N+d_2-1} {\un A} \m\Pi^{N+d_2-1}
(\1-{\un A})^{-1} = -\m\Pi_{N-d_2-1}^{N-1} + \frac 1 {v_{d_2+1}}
\le[Q,\m\Pi_{N-1}\ri](\gamma^{-1}\L)^{d_2} (\1-{\un A})^{-1}\\
&& \m{\un {\mathcal F}}^{N} =  \m{\un \Phi}_\infty\m\Pi^{N-2} {\un B} \m\Pi_{N-2}
(\1-{\un B})^{-1} =  \le[Q,\m\Pi_{N-1}\ri]\le(-
\L^t\gamma^{-1}(\1-{\un B})^{-1} + \frac 1 {v_{d_2+1}}
(\gamma^{-1}\L)^{d_2} (\1-{\un A})^{-1}\ri)
\eea

\section{Folded deformation matrices ${\ds {\mathop{{\cal U}}_{\!\!N}}\,\!_K}(x)$}

From eq. (\ref{upperfolding}) and using the folding relation
(\ref{opfolding})  we obtain the desired formula for the deformation
matrices  $\ds {{\m{\cal U}_N}\!\,_K}$
(the folded version of $\frac 1 K Q^K_{+0}$), indeed (recall $M=N-d_2-1$):
\bea
-K\,{\m{\mathcal U}_N}\!\,_K & = & \m\Pi_N^M\,Q^K_{+0} {\mathcal F}_N =
\m\Pi_N^M\,Q^K_{+0} \m\Pi^{M}\,{\mathcal F}_  =\cr
& = & \m\Pi_N^M\,Q^K_{+0}\m\Pi_{N}^M\,{\mathcal F}_  + \m\Pi_N^M\,Q^K \m\Pi^{N}\,{\mathcal F}_N = \cr
& = & \m\Pi_N^M\,Q^K_{+0} \m\Pi_{N}^M
      - \m\Pi_N^M\,Q^K\m\Pi_{N}^{N-1}
 -\m\Pi_N^M\,Q^K (\1-B)^{-1}\, (\Lambda^t \gamma^{-1})\, \le[Q, \m\Pi_{N-1}\ri] =\cr
& = & \m\Pi_N^M\,Q^K_{+0} \m\Pi_{N-1}^M
      -\m\Pi_N^M\,Q^K_{-0}\m\Pi_{N}^{N-1}  \cr
&   & - \m\Pi_N^M\,(Q^K-x^K) (\1-B)^{-1}\, (\Lambda^t \gamma^{-1})\, \le[Q, \m\Pi_{N-1}\ri] \cr
&   & - x^K\, \m\Pi_N^M\, (1-B)^{-1}\, (\Lambda^t \gamma^{-1})\, \le[Q, \m\Pi_{N-1}\ri] =\cr
& = &\m\Pi_N^M\,Q^K_{+0}\m\Pi_{N-1}^M
      - \m\Pi_N^M\,Q^K_{-0} \m\Pi_{N}^{N-1} - \m\Pi_N^M\,{Q^K-x^K\over Q-x}\, \le[Q, \m\Pi_{N-1}\ri] -
      x^K\,\m\Pi_N^{N-1}=\nonumber\\[12pt]
&=&   \m\Pi_N^M\,Q^K_{+0} \m\Pi_{N-1}^M
      -\m\Pi_N^M\,Q^K_{-0} \m\Pi_{N}^{N-1}
 -\m\Pi_N^M\,W_K(x)\, \le[Q, \m\Pi_{N-1}\ri] -
      x^K\,\m\Pi_N^{N-1} =\nonumber\\[12pt]
&=& \m\Pi_N^M\,Q^K_{+0} \m\Pi_{N-1}^M  -\m\Pi_N^M\,W_K(x)\, \le[Q, \m\Pi_{N-1}\ri] +
     \le( x^K-\frac 1 2 Q^{K}_{N,N}\ri)\,\m\Pi_N^{N-1}\ ,\label{defU}
\eea
where we have defined
\beq
W_K(x) := \frac {Q^K-x^K}{Q-x}\ .
\eeq
This formula generalizes similar formulas for the one-matrix model
appeared in \cite{BEHiso} which allowed us to perform explicit
computations linking the isomonodromic tau function and the partition
function of the model (see Conclusion).
{\bf Trace Formulas}\\
We compute the trace of the deformation matrices by noticing that the
last term in (\ref{defU}) has zero trace because $W_K(x)$ commutes
with $Q$ and because of the cyclicity of the trace
\beq
 \Tr \m\Pi^M_N W_K(x) \le[Q,\m{\Pi}^0_{n-1}\ri] =  \Tr  W_K(x)
 \le[Q,\m{\Pi}^0_{n-1}\ri]\m\Pi^M_N =  \Tr  W_K(x)
 \le[Q,\m{\Pi}^0_{n-1}\ri]= 0
\eeq
where $\Tr$ means the trace of the (finite rank) semi-infinite matrix.
Therefore
\beq
-K \tr {\m{\cal U}_N}\!\,_K(x)  =   x^K - {1\over 2}Q^K_{N,N}
 + \sum_{m=1}^{d_2} {1\over 2} Q^K_{N-m,N-m}  \ .
\eeq
Now, the deformation equations imply the deformation equation for the
normalization coefficients of the biorthogonal polynomials,
\beq
{1\over K} Q^K_{N,N}  =  - \hbar \pa_{u_K}\ln{h_N}
\eeq
which implies promptly
\beq
\tr {\m{\cal U}_N}\!\,_K(x)  =   -{1\over K}x^k
 + {\hbar\over 2} \pa_{ u_k} \ln\le({\prod_{m=1}^{d_2} h_{N-m}\over
h_N }\ri)=  -{1\over K}x^k
 + {\hbar\over 2} \pa_{ u_k} \ln\le({\mathcal Z_{N}}^2\over
\mathcal Z_{N-d_2}\mathcal Z_{N+1} \ri),\label{traceUk}
\eeq
where we have used the fact that \cite{eynardmehta}
\beq
\mathcal Z_N = C_N \prod_{j=0}^{N-1} h_j\ ,
\eeq
with $C_n$ dependent only on $N$ but independent on the $V_i$'s.

\section{Differential system ${\ds {\m{D_1}_N}(x)}$}
From the formulas for the deformation equations we can derive the
formula for the differential equation. Recall that  $P$ is of
finite band size and that we have:
\beq
P =  P_{-1} +  P_{_\geq } = P_{-} +{V'_1(Q)}_{_\geq} =\L^t\gamma
+ \sum_{K=0}^{d_1} u_{K+1}\le( Q^K_{+0} +\frac 1 2 Q^K_0\ri)\ .
\eeq
Using (\ref{eqPQV}) and (\ref{eqPQVzero}) we get:
\beq
P = \L^t\gamma +u_1\1-\sum_{K=1}^{d_1} K u_{K+1} U_K + {1\over 2}
\sum_{K=1}^{d_1} u_{K+1} Q^K_0\label{Pdeco}\ .
\eeq

Since $\ds{\m {D}_N}_1(x)$ is nothing but the folded version of $P$,
and the folding is linear, we have:
\beq
\m {D_1}_N(x)
 = \m\Pi_N^M \L^t\gamma {\mathcal F}_N +u_1\1_{d_2+1}
 - \sum_{K=1}^{d_1}  u_{K+1} K {\m{\cal
U}_N}\!\,_{K}(x) + {1\over 2} \sum_{K=1}^{d_1} u_{K+1} {\m{q}_N}^{K}_0
\eeq
where $\ds {\m\gamma_N}$ and $\ds {\m{q}_N}^{K}_0$ are the diagonal matrices:
\beq
{\m\gamma_N} = {\rm diag}\,(\gamma(N-d_2-1),\dots , \gamma(n-1))
\virg
{\m{q}_N}^{K}_0 = {\rm diag}\,(Q^K_{N-d_2,N-d_2},\dots , Q^K_{N,N})\ .
\eeq
We leave to the reader to check that the first term is
\beq
\m\Pi_N^M \L^t\gamma \m{\mathcal F}_N = \m\Pi_N^M \L^t\gamma\m\Pi_N^M + \m\Pi_N^M
\L^t\gamma \m\Pi_M\m{\mathcal F}_N=   \pmatrix{\gamma(M) & & \cr & \ddots & \cr & & \gamma(N-1) }
\pmatrix{ - {\alpha_{d_2-1}(N-1)  \over \alpha_{d_2}(N-1)} & \dots &
  {x- \alpha_{0}(N-1)  \over \alpha_{d_2}(N-1)} & - {\gamma(N-1)
    \over \alpha_{d_2}(N-1)} \cr 1 & & &  0 \cr  & \ddots & & \vdots
  \cr 0 & \dots & 1 & 0} \ ,
\eeq
where the second matrix in the product is nothing but the inverse of
the ladder matrix $\ds\m{\bf a}_{N-1}(x)$.
Using the expression (\ref{defU}) for the folded matrices $\ds{\m{\cal
    U}_N}\!\,_K$ and with simple manipulations one obtains ($M=N-d_2-1$).
\beq
\m{D_1}^N(x) = \m\Pi_N^M  V_1'(Q)_{_\geq} \m\Pi_{N-1}^{M}  - \,\m\Pi_{N}^M
W(x)\le[Q,\m\Pi_{N-1}\ri] + V_1'(x)\m\Pi^{N-1}_N + \m\Pi_N^M \L^t\gamma {\mathcal F}_N
\eeq
where $W(x)$ is the semi-infinite matrix (polynomial in $x$ and $Q$):
\beq\label{defW}
W(x) = {V'_1(Q)-V'_1(x)\over Q-x}\ .
\eeq

In a totally explicit fashion we have the expression of the matrix
$\ds{\m{D_1}_N}(x)$:
\bea
{\m {D_1}_{\!N}}(x) & = &
  \pmatrix{
V'_1(Q)_{N-d_2,N-d_2} & \dots & V'_1(Q)_{N-d_2,N-1} & 0 \cr
 0 & \ddots & \vdots & \vdots \cr
0 & 0 & V'_1(Q)_{N-1,N-1} & 0 \cr
0 & \dots & 0 & V_1'(x)
}\cr
& & +  \pmatrix{\gamma(M) & & \cr & \ddots & \cr & & \gamma(N-1) }
\pmatrix{ - {\alpha_{d_2-1}(N-1)  \over \alpha_{d_2}(N-1)} & \dots &
  {x- \alpha_{0}(N-1)  \over \alpha_{d_2}(N-1)} & - {\gamma(N-1)
    \over \alpha_{d_2}(N-1)} \cr 1 & & &  0 \cr  & \ddots & & \vdots
  \cr 0 & \dots & 1 & 0}  \cr
& & - \pmatrix{
W(x)_{M + 1,N-1} & \dots & W(x)_{M + 1,N+d_2-1}\cr
\vdots & & \vdots \cr
W(x)_{N,N-1} & \dots & W(x)_{N,N+d_2-1}
}
\Amat^N\ ,
\eea
which is the exact analogue of the very explicit formula appeared in \cite{BEHiso}.

Notice that (using (\ref{eqPQV}) and (\ref{eqPQVzero})), if $k\geq l$, we have:
\beq
V'_1(Q)_{N-k,N-l} = - P_{N-k,N-l} = \beta_{k-l}(N-l)
\eeq
which allows to rewrite:
\bea
{\m {D_1}_{\!\!N}}(x) & = &
  \pmatrix{
\beta_0(N-d_2) & \dots & \beta_{d_2-1}(N-1) & 0 \cr
 \gamma(N-d_2) & \ddots & \vdots & \vdots \cr
0 & \ddots & \beta_0(N-1) & 0 \cr
0 & \dots  & \gamma(N-1) & V_1'(x)
}+\cr
& & - {\gamma(M)\over \alpha_{d_2}(M-1)} \pmatrix{
 {\alpha_{d_2-1}(N-1)} & \dots &  \alpha_{0}(N-1)-x   &  \gamma(N-1) \cr
 0 & \dots & &  0 \cr \vdots &  & & \vdots \cr 0 & \dots &  & 0} + \cr
& & - \pmatrix{
W(x)_{N-d_2,N-1} & \dots & W(x)_{N-d_2,N+d_2-1}\cr
\vdots & & \vdots \cr
W(x)_{N,N-1} & \dots & W(x)_{N,N+d_2-1}
}
\Amat^N\ .
\eea

{\noindent \bf Trace formula}

Using the previous formulas for the deformation matrices we can
promptly obtain the trace of  the matrix $\ds {\m{D_1}_{\!\!N}}$.
\beq
\tr {\m {D_1}_{\!\!N}}(x) = V'_1(x) + \sum_{k=1}^{d_2} \beta_0(N-k) -
\gamma(M){{\alpha_{d_2-1}(N-1)  \over \alpha_{d_2}(N-1)}}
\eeq
and using eq. (\ref{eqPQV}) we have:
\beq
\alpha_{d_2}(N-1) =  \prod_{k=1}^{d_2} \gamma(N-1-k)  \,\,\, v_{d_2+1}
\eeq
and
\beq
\alpha_{d_2-1}(N-1) = \prod_{k=1}^{d_2-1} \gamma(N-1-k) \left( v_{d_2}
+ v_{d_2+1} \sum_{j=1}^{d_2} \beta_0(N-j)) \right)
\eeq
therefore:
\beq
\tr {\m {D_1}_{\!\!N}}(x) = V'_1(x) - {v_{d_2}\over v_{d_2+1}}\ .\label{TraceD}
\eeq
Note the remarkable fact that the trace of $\ds{\m {D_1}_{\!\!N}}$ does not
depend on $N$. We also point out that this result cannot be obtained
directly from the formula contained in \cite{BEH}.

\subsection{Folded deformation matrices ${\ds {\mathop{{\cal V}}_N}\!\,_J}(x)$}

We first consider the deformations w.r.t. $v_J$,  $0\leq J \leq d_2$
because they have simpler expression and allow to compute the
deformation w.r.t.  $v_{d_2+1}$  by folding of (\ref{PQrel}). The
semiinfinite deformation matrix is the finite-band size matrix
\beq
V_J=-\frac 1 J (P^J)_{-0}
\eeq
Its folded counterpart as per eq. (\ref{opfolding}) is (for brevity in
the formulas we set $M:= N-d_2-1$):
\bea
-J\,{\m{\cal V}_N}\!\,_J(x)
& = & \m\Pi_{N}^{M}\, (P^J)_{-0}\, {\mathcal F}_N  = \m\Pi_{N}^{M}\, (P^J)_{-0}\, \m\Pi_{N}\, {\mathcal F}_N   \cr
& = & \m\Pi_{N}^{M}\, (P^J)_{-0}\, \m\Pi_{N}^{M}\, {\mathcal F}_N
     +\m \Pi_{N}^{M}\, P^J\, \m\Pi_{M}\, {\mathcal F}_N   \cr
& = & \m\Pi_{N}^{M}\, (P^J)_{-0}\, \m\Pi_{N}^{M}
     +\m \Pi_{N}^{M}\, P^J\, (1-A)^{-1}\,
    \m \Pi_{M}\,A\,\m\Pi^{M}\ .\label{iits}
\eea
From the expression (\ref{Afol}) for $A$ we have the relation
\bea
 && \m\Pi_{M}A\m\Pi^{M} = -\frac 1{v_{d_2+1}}
\m\Pi_{M}(\gamma^{-1}\Lambda^t)^{d_2} (Q-x)\m\Pi^{M} = -\frac 1{v_{d_2+1}}(\gamma^{-1}\Lambda^t)^{d_2}
\m\Pi_{N-1}(Q-x)\m\Pi^{M}=\cr
&&= -\frac 1{v_{d_2+1}}(\gamma^{-1}\Lambda^t)^{d_2}
(Q-x)\m\Pi_{N-1}\m\Pi^{M} + \frac
1{v_{d_2+1}}(\gamma^{-1}\Lambda^t)^{d_2} \le[Q,\m\Pi_{N-1}\ri]=\cr
&&= -(1-A)\m\Pi^{M}_{N-1} + \frac
1{v_{d_2+1}}(\gamma^{-1}\Lambda^t)^{d_2} \le[Q,\m\Pi_{N-1}\ri]\ .\label{iits1}
\eea
Plugging (\ref{iits1}) into (\ref{iits}) we obtain
\bea
-J\,{\m{\cal V}_N}\!\,_J(x)
 & = & \m\Pi_N^{M} P^J_{-0}\m\Pi_N^{M} -
\m\Pi_{N}^{M} P^J \m\Pi_{N-1}^{M} +
\frac 1{v_{d_2+1}} \m\Pi_N^{M} P^J(1-A)^{-1}  (\gamma^{-1}
\Lambda)^{d_2}\,\le [Q, \m\Pi_{N-1}\ri]\ .
\eea
The last deformation matrix is obtained from eq. (\ref{PQrel}) by also
recalling that $\m\Pi^{d_2+1} Q {\mathcal F}_N = x \m\Pi^{d_2+1} {\mathcal F}_N$ to be
\beq
-(d_2+1)v_{d_2+1} {\m{\mathcal V}_N}\!\,_{d_2+1} = x\m{D_1}_{\!\!N}(x)+ \sum_{J=1}^{d_2}
J\,v_{J}{\m{\mathcal V}_N}\!\,_J +  \sum_{K=1}^{d_1+1}
K\,u_{K}{\m{\mathcal U}_N}\!\,_K - \hbar {\rm diag}(N-d_2,\dots,N)\label{vd21}
\eeq
\subsubsection{Trace}

We will compute first the traces of the deformation matrices for
$J\leq d_2$, whereas the case $J=d_2+1$ will be obtained by tracing
the identity (\ref{vd21}). Hereafter the notation $\tr$ denotes the
finite dimensional trace for the folded matrices (of dimension
$(d_2+1)\times(d_2+1)$) while $\Tr$ denotes the trace of a
semiinfinite matrix. Bearing this in mind we compute
\bea
-J\,\tr {\m{\cal V}_N}\!\,_J(x)
  =  {1\over 2}\Tr\le( P^J\m \Pi_{N}^{N-d_2-1}\ri) - \Tr \le(P^J
\m\Pi_{N-1}^{N-d_2-1}\ri)\,  + {1\over v_{d_2+1}}\, \Tr\le( P^J\,
(1-A)^{-1}\, (\gamma^{-1} \Lambda)^{d_2}\, \le[Q, \m\Pi_{N-1}\ri
]\ri)=\cr
= \frac 1 2 P^J_{N,N} -\frac 1 2  \sum_{j=M}^{N-1}P^k_{j,j} + {1\over v_{d_2+1}}\, \Tr\le( P^J\,
(1-A)^{-1}\, (\gamma^{-1} \Lambda)^{d_2}\, \le[Q, \m\Pi_{N-1}\ri
]\ri)
\eea
Let us focus our attention on the last term.
First, notice that:
\bea
&   & {1\over v_{d_2+1}}\, \Tr P^J\, (1-A)^{-1}\, (\gamma^{-1} \Lambda)^{d_2}\, [Q, \Pi_{N-1}]\, \cr
& = & {1\over v_{d_2+1}}\, \Tr P^J\, (1-\td{A})^{-1}\, (\gamma^{-1} \Lambda)^{d_2}\, [V'_2(P)-x, \Pi_{N-1}]\, \cr
\eea
where $\td{A}$ is the following strictly upper triangular matrix
\beq
\td{A} :=1-(\gamma^{-1} \Lambda)^{d_2}(V'_2(P)-x)\label{Atilde}\ .
\eeq
Indeed, for $J\leq d_2$, the difference is the trace of a strictly upper triangular matrix.
Then we have:
\beq
\Tr P^J\, (\1-\td{A})^{-1}\, (\gamma^{-1} \Lambda)^{d_2}\, \le[V'_2(P)-x, \m\Pi_{N-1}\ri]
= \Tr P^J\, \le[ (\1-\td{A})^{-1}\, (\gamma^{-1} \Lambda)^{d_2}, V'_2(P)-x\ri]\, \m\Pi_{N-1}
\eeq

Let us denote by  $y_j = y_j(x)$, $j=1,\dots,d_2$ the solutions of algebraic
equation $V'_2(y)=x$ and define the following polynomials in $y$:
\beq
S_j(y) := \frac {V'_2(y)-x}{ y-y_j}
\eeq
We have in particular $S_j(y_j)=V''_2(y_j)$.

We also define the following strictly upper triangular matrices:
\beq
R_j := \1-(\gamma^{-1} \Lambda)\,(P-y_j)
\eeq
We are now going to use the following identity, whose proof can be
found in appendix \ref{appendix}:
\beq
(\1-\td{A})^{-1}(\gamma^{-1} \Lambda)^{d_2} = \sum_{j=1}^{d_2}
{1 \over V''_2(y_j)}\, (\1-R_j)^{-1} \gamma^{-1} \Lambda\label{inappendix}
\eeq
Using (\ref{inappendix}) we have:
\bea
&   & {1\over v_{d_2+1}}\, \Tr P^J\, \le[ (\1-\td{A})^{-1}\, (\gamma^{-1}
  \Lambda)^{d_2}, V'_2(P)-x\ri]\, \m\Pi_{N-1} \cr 
& = & \sum_{j=1}^{d_2} {1\over V''_2(y_j)}\, \Tr P^J\, \le[ (\1-R_j)^{-1}\,
   \gamma^{-1} \Lambda, V'_2(P)-x \ri]\, \m\Pi_{N-1}\, \cr
& = & \sum_{j=1}^{d_2} {1\over V''_2(y_j)}\, \Tr P^J\, (\1-R_j)^{-1}\,
 \gamma^{-1} \Lambda (V'_2(P)-x)\, \m\Pi_{N-1}  \cr
&   & - \sum_{j=1}^{d_2}
       {1\over V''_2(y_j)}\, \Tr P^J\, (V'_2(P)-x)\,
 (\1-R_j)^{-1}\, \gamma^{-1} \Lambda \m\Pi_{N-1}\,=  \cr
& = & \sum_{j=1}^{d_2} {1\over V''_2(y_j)}\, \Tr P^J\,
\overbrace{(\1-R_j)^{-1}\,
\gamma^{-1} \Lambda (P-y_j)}^{=\1}{V'_2(P)-x\over P-y_j}\, \m\Pi_{N-1}\,+ \cr
&   & - \sum_{j=1}^{d_2} {1\over V''_2(y_j)}\, \Tr P^J\,
      {V'_2(P)-x\over P-y_j}\,( \Lambda^t\gamma \gamma^{-1}
      \Lambda+\m\Pi_0)(P-y_j)  (\1-R_j)^{-1}\, \gamma^{-1} \Lambda \m\Pi_{N-1}\, \cr
& = & \sum_{j=1}^{d_2} {1\over V''_2(y_j)}\, \Tr P^J\, S_j(P)\, \m\Pi_{N-1}\,
      - \sum_{j=1}^{d_2} {1\over V''_2(y_j)}\, \Tr P^J\, S_j(P)\, \Lambda^t \Lambda \m\Pi_{N-1}\, \cr
&   & - \sum_{j=1}^{d_2} {1\over V''_2(y_j)}\, \Tr P^J\,
      S_j(P)\,\m\Pi_0(P-y_j)  (1-R_j)^{-1}\, \gamma^{-1} \Lambda \m\Pi_{N-1}\,
      \cr
& = & \sum_{j=1}^{d_2} {1\over V''_2(y_j)}\, \Tr P^J\, S_j(P)\, \m\Pi_0
      - \sum_{j=1}^{d_2} {1\over V''_2(y_j)}\, \Tr P^J\,
      S_j(P)\,\m\Pi_0(P-y_j)  (\1-R_j)^{-1}\, \gamma^{-1} \Lambda\m \Pi_{N-1}\, \cr
\eea
Notice that if $N>2d_2$ the $\Pi_{N-1}$ in the last trace is irrelevant and therefore that trace is independent of $N$.
Assuming thus $N>2d_2$ we obtain can carry on with our computation
\bea
&   & {1\over v_{d_2+1}}\, \Tr P^J\, (1-A)^{-1}\, (\gamma^{-1} \Lambda)^{d_2}\, [Q, \Pi_{N-1}]\, \cr
& = & \sum_{j=1}^{d_2} {1\over V''_2(y_j)}\, \Tr P^J\, S_j(P)\, \m\Pi_0
      - \sum_{j=1}^{d_2} {1\over V''_2(y_j)}\, \Tr \m\Pi_0(P-y_j)  (1-R_j)^{-1}\, \gamma^{-1} \Lambda\, P^J\, S_j(P) \cr
& = & \sum_{j=1}^{d_2} {1\over V''_2(y_j)}\, \Tr P^J\, S_j(P)\, \m\Pi_0
      - \sum_{j=1}^{d_2} {y_j^k\over V''_2(y_j)}\, \Tr \m\Pi_0(P-y_j)  (1-R_j)^{-1}\, \gamma^{-1} \Lambda\, S_j(P) \cr
&   & - \sum_{j=1}^{d_2} {1\over V''_2(y_j)}\, \Tr \m\Pi_0(P-y_j)  (1-R_j)^{-1}\, \gamma^{-1} \Lambda\, (P^J-y_j^J)\, S_j(P) \cr
& = & \sum_{j=1}^{d_2} {1\over V''_2(y_j)}\, \Tr P^J\, S_j(P)\, \m\Pi_0
      - \sum_{j=1}^{d_2} {y_j^J\over V''_2(y_j)}\, \Tr \m\Pi_0(P-y_j)  (1-R_j)^{-1}\, \gamma^{-1} \Lambda\, S_j(P) \cr
&   & - \sum_{j=1}^{d_2} {1\over V''_2(y_j)}\, \Tr \m\Pi_0(P-y_j)  (1-R_j)^{-1}\, \gamma^{-1} \Lambda\, (P-y_j)
             {P^J-y_j^J\over P-y_j}\, S_j(P) \cr
& = & \sum_{j=1}^{d_2} {1\over V''_2(y_j)}\, \Tr P^J\, S_j(P)\, \m\Pi_0
      - \sum_{j=1}^{d_2} {y_j^J\over V''_2(y_j)}\, \Tr \m\Pi_0(P-y_j)  (1-R_j)^{-1}\, \gamma^{-1} \Lambda\, (S_j(P)-S_j(y_j)) \cr
&   & - \sum_{j=1}^{d_2} y_j^J\, \Tr (P-y_j)  (1-R_j)^{-1}\, \gamma^{-1} \Lambda\, \m\Pi_0
      - \sum_{j=1}^{d_2} {1\over V''_2(y_j)}\, \Tr \m\Pi_0(P-y_j)
      {P^J-y_j^J\over P-y_j}\, S_j(P)\label{1212}\ .
\eea
Recalling that $\ds\Lambda \m\Pi_0 =0$ we continue
\bea
\hbox{(\ref{1212})}
&=   & \sum_{j=1}^{d_2} {1\over V''_2(y_j)}\, \Tr P^J\, S_j(P)\, \m\Pi_0
      - \sum_{j=1}^{d_2} {1\over V''_2(y_j)}\, \Tr \m\Pi_0 (P^J-y_j^J)\, S_j(P) \cr
&   & - \sum_{j=1}^{d_2} {y_j^J\over V''_2(y_j)}\, \Tr \m\Pi_0(P-y_j)  (1-R_j)^{-1}\, \gamma^{-1} \Lambda\, (P-y_j)
{S_j(P)-S_j(y_j)\over P-y_j} \cr
& = & \sum_{j=1}^{d_2} {y_j^J\over V''_2(y_j)}\, \Tr S_j(P)\, \m\Pi_0
      - \sum_{j=1}^{d_2} {y_j^J\over V''_2(y_j)}\, \Tr \m\Pi_0(P-y_j) {S_j(P)-S_j(y_j)\over P-y_j} \cr
& = & \sum_{j=1}^{d_2} {y_j^J\over V''_2(y_j)}\, \Tr S_j(P)\, \m\Pi_0
      - \sum_{j=1}^{d_2} {y_j^J\over V''_2(y_j)}\, \Tr \m\Pi_0(S_j(P)-S_j(y_j)) \cr
& = & \sum_{j=1}^{d_2} y_j^J\, \Tr \m\Pi_0
 =  \sum_{j=1}^{d_2} y_j^J \label{final1}
\eea
With eq. (\ref{final1}) we can finally state
\beq
\encadremath{
-J\,\tr {\m{\cal V}_N}\!\,_J(x)
  =   \frac 1 2 P^J_{N,N} -\frac 1 2  \sum_{j=M}^{N-1}P^J_{j,j}
    + \sum_{j=1}^{d_2} y_j^J
}\label{tracesvj}
\eeq
Using eq. (\ref{tracesvj}) and tracing (\ref{vd21}) one can obtain the
following formulas for the traces
\bea
\tr {\m{\mathcal V}_N}\!\,_J(x) = -\frac \hbar 2 \pa_{v_J}
\ln\le(\frac {\prod_{j=1}^{d_2} h_{N-j}}{h_N} \ri) +\oint {\rm d}y\,
\frac {y^J}{J} \frac {V_2''(y)}{V_2'(y)-x}\ ,\ \ \
J=1,\dots,d_2\nonumber \\
\tr {\m{\mathcal V}_N}\!\,_{d_2+1}(x) = -\frac \hbar 2 \pa_{v_{d_2+1}}
\ln\le(\frac {\prod_{j=1}^{d_2} h_{N-j}}{h_N} \ri) +\oint {\rm d}y\,
\frac {y^{d_2+1}}{d_2+1} \frac {V_2''(y)}{V_2'(y)-x}-\hbar \le(N -
\frac{d_2}2 \ri)\frac 1{v_{d_2+1}} \ .\label{guess2}
\eea

\subsection{Determinant of the fundamental system}

It was proven in \cite{BEH} that the following
overdetermined set of differential equations for a
$(d_2+1)\times(d_2+1)$ matrix $\ds \m{\bf \Psi}_N$ is Frobenius compatible:
\bea
&& -\hbar \pa_x \m{\bf \Psi}_N(x) = {\m{D_1}_{\!\!N}}(x)\m{\bf \Psi}_N(x)\label{11}\\
&& \hbar \pa_{u_K}\m{\bf \Psi}_N(x) = {\m{\mathcal U}_N}\!\,_K(x) \m{\bf
\Psi}_N(x)\label{12}\\
&& -\hbar \pa_{v_J}\m{\bf \Psi}_N(x) = {\m{\mathcal V}_N}\!\,_J(x) \m{\bf
\Psi}_N(x)\label{13}\\
&& \m{\bf a}_N(x)\m{\bf \Psi}_N(x) = \m{\bf \Psi}_{N+1}(x)\label{14}
\eea
The biorthogonal polynomials constitute one column of $\ds \m{\bf \Psi}_N(x)$, while the other $d_2$ can
be also explicitly obtained as in \cite{Asym}.

From (\ref{guess2})  it follows that the dependence on the coefficients
of $V_2$ of the determinant of the fundamental system is
\bea
\hbar{\rm d}_{\bf v}\ln\det\le(\m{\bf
  \Psi}_N(x)\ri) :=  -\sum_{J=1}^{d_2+1} {\rm d}v_J \hbar\pa_{v_J} \ln\det\le(\m{\bf
  \Psi}_N(x)\ri) =  \sum_{J=1}^{d_2+1} {\rm d}v_J \tr\le(
{\m{\mathcal V}_N}\!\,_J(x)\ri) = \\
=-\hbar {\rm d}_{\bf v} \ln\le(\sqrt{\frac
  {\prod_{j=1}^{d_2}h_{N-j}}{h_N}} \ri) +  \sum_{J=1}^{d_2+1} {\rm
  d}v_J \oint {\rm d}y \frac {y^J}{J} \frac {V_2''(y)}{V_2'(y)-x} -\hbar
\le(N-\frac {d_2}2\ri){\rm d}\ln(v_{d_2+1})
\eea
Now one can check directly that
\beq
\oint {\rm d} y\, \frac {y^J}{J} \frac {V_2''(y)}{V_2'(y)-x}  = \pa_{v_J}\le(
 \oint {\rm d} y\, \le(V_2(y)-xy\ri) \frac {V_2''(y)}{V_2'(y)-x} \ri)
\eeq
Therefore, as a consequence of eqs. (\ref{TraceD}, \ref{guess2},
\ref{traceUk}) we find  that any matrix fundamental
solution to the compatible system of equations
satisfies the equations
\bea
&& \hbar \pa_x \ln\le( \det \m{\bf \Psi}_N(x)\ri) = -V_1'(x) + \frac
{v_{d_2}}{v_{d_2+1}}\\
&& \hbar{\rm d}_{\bf u}\ln\le( \det \m{\bf
  \Psi}_N(x)\ri) =  {\rm d}_{\bf u}\le(-V_1(x)
 + {\hbar\over 2}  \ln\le({\prod_{m=1}^{d_2} h_{N-m}\over
h_N }\ri)\ri)
\\
&& \hbar{\rm d}_{\bf v}\ln\le( \det \m{\bf
  \Psi}_N(x)\ri) = -\hbar {\rm d}_{\bf v}
\le[
 \ln\le({v_{d_2+1}}^{\frac {d_2} 2-N}\sqrt{\frac
  {\prod_{m=1}^{d_2}h_{N-m}}{h_N}} \ri) +
 \oint {\rm d} y\, \le(V_2(y)-xy\ri) \frac {V_2''(y)}{V_2'(y)-x}
\ri]\\
&& \det\le( \m{\bf \Psi}_{N+1}(x) \m{\bf \Psi}_N(x)^{-1} \ri) =
\det(\m{\bf a}_N(x))=
v_{d_2+1}(-)^{d_2+1} \sqrt {\frac {{h_N}^2 h_{N-d_2-1}}{h_N+1}}.
\eea

and hence we have finally the complete formula
\beq
\det\le[\m{\bf \Psi}_N(x) \ri]  =  (v_{d_2+1})^{N-\frac {d_2}2}
(-1)^{N(d_2+1)}\sqrt {\prod_{m=1}^{d_2} h_{N-m}\over
h_N } \exp\le[-\frac 1 \hbar \le( V_1(x) +\oint {\rm d}y (V_2(y)-xy)
  \frac {V_2''(y)}{V_2'(y)-x}\ri)\ri] \ .
\eeq
Note that the contour integral just returns the sum of the critical
values of $V_2(y)-xy$.

We know from \cite{BEH} that  a joint solution of the dual system of
overdetermined equations for the dual window denoted by
$\ds\m{\bf\un\Phi}^N(x)$ has the property that
\beq
\m{\bf \un\Phi}^N(x)\Amat^N \m{\bf \Psi}_N(x) = C\ ,
\eeq
where $C$ is an invertible matrix which does not depend on any of the
potentials or $n$ or $\hbar$ or $x$ and can be conveniently be
normalized to unity (see \cite{Asym} for an explicit construction).
Therefore
\bea
\det(\m{\bf\un\Phi}^N(x))  &=& \det(\Amat^N)^{-1}\det( \m{\bf
  \Psi}_N(x))^{-1}
\eea

\section{Dual folded system}
For completeness we add the formulas for the relevant folded
operators for the dual window
$\ds \m \Phi^N =(\phi_{N-1},\dots,\phi_{N+d_2-1})$. 
Since the steps are essentially
the same we give only the results (here we set $L=N+d_2-1$)
\bea
&&\hspace{-2truecm} K{\m{\un U}^N}_K = \m\Pi_{L}^{N-1} Q^{K}_{+0}\m\Pi_{L}^{N-1} -
\le[Q,\Pi_{N-1}\ri] W_K(x)\m\Pi^{N-1}_{L}  + \le(x^K-\frac 1 2
Q^{K}_{N-1,N-1}\ri) \m\Pi^{N-1}_N\\
&&\hspace{-2truecm} J{\m{\un V}^N}_J = \m\Pi^{N-1}_L P^J_{-0}\m\Pi^{N-1}_{L} +
\m\Pi_L \un A \m\Pi^L (\1-\un A)^{-1}  P^{J}\m\Pi^{N-1}_{L}\\
&&\hspace{-2truecm}{\m{\un{D}}^N}_1 (x)=
\pmatrix{ V'_1(x)  & 0 & \dots & 0 \cr \gamma(N-1) & \beta_0(N) &
\dots &  \beta_{d_2-1}(L) \cr 0 & \ddots & \ddots & \vdots \cr 0
&\dots         & \gamma(L-1) & \beta_0(L) \cr }+\\
&&\hspace{-1.5truecm}-\frac {\gamma(L)}{\alpha_{d_2}(L+1)}
\pmatrix{ 0 &\dots  & 0 & {\gamma(N-1)} \cr 0 &   & &{\alpha_0(N)-x}
\cr 0 &\ddots   &        & \vdots\cr 0 &\dots &0 &
{\alpha_{d_2-1}(L)}} -\Amat^N \pmatrix{ W(x)_{N-d_2,N-1}
& \dots &W(x)_{N-d_2,N+d_2-1}\cr \vdots & & \vdots \cr W(x)_{N,N-1} &
\dots & W(x)_{N,N+d_2-1}  }
 \eea

\section{Conclusion}
We have given the most explicit construction for the matrices
describing the differential-deformation-difference folded system.
This has allowed us to determine explicitly the determinant of the
fundamental solution of the system in terms of the partition function
of the model and the two potentials.
The final expression is quite simple in comparison with the complexity
of the computation, expecially for the traces of the deformation
matrices $\ds\m{{\cal V}_J}_N$.\par
It is our hope (in fact it is our plan) that these formula be used to
relate explicitly the partition functio of the two--matrix model to a
(suitably defined) isomonodromic tau--function.
Indeed the system (\ref{11},\ref{12},\ref{13},\ref{14}) can (and should) be regarded as
a monodromy--preserving set of infinitesimal (\ref{12}, \ref{13}) and
finite (\ref{14}) deformation equation for the ODE (\ref{11}) which
has an irregular, degenerate singularity at infinity. As such it is
envisionable that one can define the associated notion of
tau--function  although there are technical difficulties due to the degeneracy
at infinity.
In the similar case of the ODE+PDE+$\Delta$E for the orthogonal
polynomials in the one--matrix model a similar approach has given
very satisfactory results \cite{BEHiso, BlIt}.


\setcounter{section}{0}
\appendix
\section{Proof of formula (\ref{inappendix})}
\label{appendix}
We need to prove formula (\ref{inappendix})
\beq
(\1-\td{A})^{-1}(\gamma^{-1} \Lambda)^{d_2} =
\sum_{j=1}^{d_2} {1 \over V''_2(y_j)}\, (\1-R_j)^{-1}
\gamma^{-1} \Lambda\ ,
\eeq
which can be written more transparently as
\beq
\le((\gamma^{-1} \Lambda)^{d_2}(V'_2(P)-x)\ri)^{-1}(\gamma^{-1} \Lambda)^{d_2} = \sum_{j=1}^{d_2}
\frac 1 {V_2''(y_j)} \le(\gamma^{-1}\Lambda(P-y_j)\ri)^{-1}\gamma^{-1}\Lambda\ .
\eeq
We remark for the better understanding of the reader that if the
matrices $\Lambda, \ (V_2'(P)-x)$ were invertible this would amount simply to the
partial fraction expansion  of $1/(V_2'(P)-x)$.
Multiplying both sides by the invertible matrix $(\1-\td{A})$ on the left and recalling the
definition (\ref{Atilde}) of $\td A$
\bea
(\gamma^{-1} \Lambda)^{d_2}
& = & (\gamma^{-1} \Lambda)^{d_2} \le(\sum_{j=1}^{d_2} {1\over
  V''_2(y_j)}\, (V'_2(P)-x)(\1-R_j)^{-1} \gamma^{-1} \Lambda \ri)\cr
& = & (\gamma^{-1} \Lambda)^{d_2} \le( \sum_{j=1}^{d_2} {1\over
  V''_2(y_j)}\,  {V'_2(P)-x\over P-y_j}(P-y_j)(\1-R_j)^{-1} \gamma^{-1} \Lambda \ri)\cr
& = & (\gamma^{-1} \Lambda)^{d_2} \le(\sum_{j=1}^{d_2} {1\over V''_2(y_j)}\, {V'_2(P)-x\over
	 P-y_j}(\m\Pi_0+\Lambda^t\gamma \gamma^{-1}
       \Lambda)(P-y_j)(\1-R_j)^{-1} \gamma^{-1} \Lambda \ri)\cr
\eea
Noticing now  that $\ds (\gamma^{-1} \Lambda)^{d_2} {V'_2(P)-x\over P-y_j}
\m\Pi_0=0$ we continue
\bea
(\gamma^{-1} \Lambda)^{d_2}
& = &(\gamma^{-1} \Lambda)^{d_2} \le (
      \sum_{j=1}^{d_2} {1\over V''_2(y_j)}\, {V'_2(P)-x\over P-y_j} \Lambda^t\gamma \gamma^{-1} \Lambda \ri) \cr
& = & (\gamma^{-1} \Lambda)^{d_2} \le(\sum_{j=1}^{d_2} {1\over
	 V''_2(y_j)}\, {V'_2(P)-x\over P-y_j} \le(\1-\m\Pi_0\ri) \ri) \cr
& = & (\gamma^{-1} \Lambda)^{d_2} \le(\1-\m\Pi_0\ri)
 =  (\gamma^{-1} \Lambda)^{d_2}\ ,
\eea
where we have used  that $\ds(\gamma^{-1} \Lambda)^{d_2}\m\Pi_0=0$. The
identity is thus proved.


\begin{thebibliography}{99}

\bibitem{AvM1} M. Adler and P. Van Moerbeke, ``String-orthogonal polynomials,
string equations and 2-Toda symmetries'', {\it Comm. Pure and Appl. Math. J.},
{\bf 50} 241-290 (1997).

\bibitem{AvM2} M. Adler and P. Van Moerbeke, ``The Spectrum of Coupled Random
Matrices'', {\it Ann. Math.} {\bf 149}, 921--976 (1999).

\bibitem{bauldry} W. Bauldry, ``Estimates of asymmetric Freud
Polynomials on the real line'', {\em J. Approx. Theory}
{\bf 63}, 225-237 (1990).

\bibitem{Asym} M. Bertola, B. Eynard, J. Harnad,
``Differential systems for biorthogonal polynomials appearing in
2-matrix models and the associated Riemann-Hilbert problem'',
CRM-2852, Saclay-T02/097, nlin.SI/0208002, to appear in
Commun. Math. Phys. in press (2002).

\bibitem{BEH} M. Bertola, B. Eynard, J. Harnad, ``Duality,
Biorthogonal Polynomials and Multi--Matrix Models'',
Commun. Math. Phys. {\bf 229} 73--120 (2002).

\bibitem{Needs} M. Bertola, B. Eynard, J. Harnad, ``Duality of
spectral curves arising in two-matrix models'',
Theor. Math. Phys. {\bf 134} no. 1, 27--32 (2003).

\bibitem{Marco} M. Bertola, ``Bilinear semi--classical moment
functionals and their integral representation'', J. App. Th. {\bf 121}
71--99 (2003).

\bibitem{BEHiso} M. Bertola, B. Eynard, J. Harnad,
``Partition functions for matrix models and isomonodromic tau functions'' Accepted in J. Phys. A.

\bibitem {BlIt} P. Bleher, A. Its, ``Semiclassical asymptotics of
orthogonal polynomials, Riemann-Hilbert problem,
and universality in the matrix model'' {\em Ann. of Math.}
 (2) {\bf 150}, no. 1, 185--266 (1999).

\bibitem{bonan} S.S. Bonan, D.S. Clark, ``Estimates of the Hermite and
the Freud polynomials'', {\em  J. Approx. Theory} {\bf 63}, 210-224 (1990).



\bibitem{KazakoVDK} J.M. Daul, V. Kazakov, I.K. Kostov, ``Rational Theories of
2D Gravity from the Two-Matrix Model'', {\em Nucl. Phys.} {\bf B409}, 311-338
(1993), hep-th/9303093.

\bibitem{dkmvz} P. Deift, T. Kriecherbauer, K. T. R. McLaughlin,
S. Venakides, Z. Zhou, ``Uniform asymptotics for polynomials
orthogonal with respect to varying exponential weights and
applications to universality questions in random matrix theory'', {\it
Commun. Pure Appl. Math.} {\bf 52}, 1335--1425 (1999).

\bibitem{dkmvz2} P. Deift, T. Kriecherbauer, K. T. R. McLaughlin,
S. Venakides, Z. Zhou, ``Strong asymptotics of orthogonal polynomials
with respect to exponential weights'', {\it Commun. Pure Appl. Math.}
{\bf 52}, 1491--1552, (1999).

\bibitem{ZJDFG} P. Di Francesco, P. Ginsparg, J. Zinn-Justin, ``2D Gravity and
Random Matrices'', {\em Phys. Rep.} {\bf 254}, 1 (1995).

\bibitem{McLaughlin} N. M. Ercolani and K. T.-R. McLaughlin ``Asymptotics and
integrable structures for biorthogonal polynomials associated to a random
two-matrix model'', {\em Physica D},  {\bf 152-153}, 232-268 (2001).

\bibitem{eynardmehta} B. Eynard, M.L. Mehta, ``Matrices coupled in a chain:
eigenvalue correlations'', {\em J. Phys. A: Math. Gen.} {\bf 31}, 4449 (1998),
cond-mat/9710230.


\bibitem{FIK} A. Fokas, A. Its, A. Kitaev, ``The isomonodromy approach
to matrix models in 2D quantum gravity'', {\em Commun. Math. Phys.}
{\bf 147}, 395--430 (1992).


\bibitem{JMU} M. Jimbo, T. Miwa and K. Ueno, ``Monodromy Preserving
Deformation of Linear Ordinary Differential Equations with Rational
Coeefficients I.'', {\it Physica} {\bf 2D}, 306-352 (1981).

\bibitem{JM}  M. Jimbo and T. Miwa, ``Monodromy Preserving Deformation of
Linear Ordinary Differential Equations with Rational Coeefficients II, III'',
{\it Physica} {\bf 2D}, 407-448 (1981); {\it ibid.}, {\bf 4D}, 26--46 (1981).

\bibitem{kapa} A. A. Kapaev, ``The Riemann--Hilbert problem for the
bi-orthogonal polynomials'', J.Phys. A {\bf 36} (2003) 4629--4640.


\bibitem{Kazakov} V.A. Kazakov, ``Ising model on a dynamical planar random
lattice: exact solution'', {\em Phys Lett.} {\bf A119}, 140-144 (1986).


\bibitem{UT} K. Ueno and K. Takasaki, ``Toda Lattice Hierarchy'',
{\it Adv. Studies Pure Math.} {\bf 4}, 1--95 (1984).

\end{thebibliography}
\end{document}